\documentstyle[aps,epsfig,preprint]{revtex}
\def\be{\begin{eqnarray}}
\def\ee{\end{eqnarray}}
\newcommand{\bp}{\bbox{p}}
\newcommand{\bP}{\bbox{P}}

\newcommand{\br}{\bbox{r}}
\newcommand{\nablavec}{\bbox{\nabla}}
\newcommand{\Deltavec}{\bbox{\Delta}}

\newcommand{\bR}{\bbox{R}}
\newcommand{\Tr}{\rm Tr}

\begin{document}
\tighten
\title{Spatially inhomogeneous condensate in asymmetric nuclear matter}
\author{A. Sedrakian}
\address{
         Kernfysisch Versneller Instituut, NL-9747 AA Groningen,
         The Netherlands
        }
\date{\today}

\maketitle
\begin{abstract}
We study the isospin singlet pairing in asymmetric nuclear matter
with nonzero total momentum of the condensate Cooper pairs.
The quasiparticle excitation spectrum is fourfold split compared
to the usual BCS spectrum of the symmetric, homogeneous matter.
A twofold splitting of the spectrum into separate branches is due
to the finite momentum of the condensate, the isospin asymmetry,
or the finite quasiparticle lifetime. The coupling of the isospin
singlet and triplet paired states leads to further twofold splitting
of each of these branches. We solve the gap equation numerically
in the isospin singlet channel in the case where the pairing in
the isospin triplet channel is neglected and find nontrivial
solutions with finite total momentum of the pairs. The corresponding
phase assumes a periodic spatial structure which carries a isospin
density wave at constant total number of particles. The phase
transition from the BCS to the inhomogeneous superconducting phase is
found to be  first order and occurs when the density asymmetry is
increased above 0.25. The transition from the inhomogeneous
superconducting to the unpaired normal state is second order. The
maximal values of the critical total momentum  (in units of the Fermi
momentum) and the critical density asymmetry at which condensate
disappears are $P_c/p_F = 0.3$ and $\alpha_c = 0.41$. The possible
spatial forms of the ground state of the inhomogeneous superconducting
phase are briefly discussed.
\end{abstract}

\section{Introduction}

The theory of fermion pairing when the fermions, which are bound in
Cooper pairs, lie on different Fermi surfaces was addressed
in the early 1960s in the context of metallic superconducting
alloys containing paramagnetic impurities\cite{AG,CLOGSTON,CHANDRA,GR}.
Recently, the theory has received renewed attention in
several contexts, including the pairing in asymmetric nuclear
matter\cite{ALM_ETAL1,BALDO,ALM_ETAL2,SAL,SL,ROEPKE_ETAL,AKHIEZER}
and color superconductivity in flavor-asymmetric high-density QCD\cite{ALFORD,BEDAQUE}.

This paper elaborates on an earlier suggestion\cite{SAL} that the
ground state of the superconducting asymmetric nuclear matter
at large asymmetries corresponds to a pair condensate
with nonzero total momentum of the Cooper pairs.
The argument is based on the observation by Larkin and Ovchinnikov\cite{LARKIN}
and Fulde and Ferrell \cite{FULDE} who first showed, in the context of
the metallic superconductors, that the Bardeen-Cooper-Schrieffer (BCS)
equations admit solutions with nonzero total momentum of Cooper pairs.
In the configuration space such a condensate forms a  periodic lattice
with finite shear modulus. The resulting spatially inhomogeneous
superconducting state is called the Larkin-Ovchinnikov-Fulde-Ferrell
(LOFF) phase.

The occurrence of pair correlations crucially depends upon
the overlap between the neutron and proton Fermi surfaces;
the pairing gap is largest in the isospin symmetric case
and is suppressed as the system is driven out of the
symmetric state. The thermal smearing of the Fermi surfaces
promotes the pairing, but, however is ineffective when the separation
between the surfaces is large compared to the temperature.
If the total momentum of the Cooper pairs is zero, the Fermi
surfaces (for homogeneous systems) are located on
concentric spheres. If, however, a Cooper pair moves with a
finite momentum, the centers of the spheres are shifted. This allows for
the pairing, even for vastly different radii of the spheres,
since now the nonconcentric spheres may intersect. The overlap regions
then provide the kinematical phase space for pairing phenomena to occur.

The microscopic calculations, based on the BCS
theory for the bulk nuclear matter show that the isospin-symmetric
matter supports Cooper-type pair correlations in the $^3S_1$-$^3D_1$
partial-wave channel due to the tensor component of the nuclear force.
The energy gap is of the order of 10 MeV at the empirical saturation
density (Refs. \cite{ALM_ETAL1,BALDO,ALM_ETAL2,SAL,SL,ROEPKE_ETAL,AKHIEZER}
and references therein) if one
assumes that the effective pairing interaction can be approximated
by the bare interaction and if the renormalization of the single particle
spectrum reduces the density of the state at the Fermi
surface by a factor of 2.

There is little evidence for large gap isospin singlet pairing
in ordinary nuclei\cite{GOODMAN}, which is evidently
suppressed due to the spin-orbit splitting\cite{LANGANKE}.
The laboratory data do not exclude the possibility that the
bulk nuclear matter, as encountered in the
supernovas and neutron stars, may support large gap pairing in the isospin
singlet channel. In the model of ``nucleon stars''\cite{BROWN} the kaon
condensation implies nearly isospin symmetric matter, in which case the
isospin singlet pairing can play a major role in determining the cooling
and rotation dynamics of such objects. Note, however, that in the models
without meson condensates the proton concentration in
supernova and neutron star matter is of the order of $5\%-30\%$ and these
asymmetries are too large to allow for neutron-proton pairing.
In the high-density regime, the hyperon-rich neutron star matter may
be much more symmetric than at the densities around the saturation
density\cite{GLENDENNING}
and therefore can support neutron-proton pairing, most likely
due to the attractive $^3D_2$ partial-wave interaction\cite{ALM_ETAL2}.
If the nucleon-hyperon and hyperon-hyperon interactions
are attractive, the pairing among fermions lying on different Fermi
surfaces, and in particular the formation of the LOFF phase,
could be a realistic possibility in hyperon rich matter.
Finally, in superstrong magnetic fields of highly magnetized neutrons stars
(magnetars) the Pauli paramagnetism will cause a splitting in the
Fermi energies of the spin-up and spin-down fermions; in this case
the pairing in the isospin triplet channel among the $S$-wave paired
neutrons should exhibit the properties of the
LOFF phase.\footnote{This problem is currently under study.}

Before turning to the main body of the paper, we draw
the reader's attention to  Ref.~\cite{QCDLOFF} who considered
the color superconducting LOFF phase in the context of high-density
QCD and  Ref.~\cite{QCDCC} who considered the
finite momentum pairing between quarks in the particle-hole channel
(chiral condensate). Our work, to some extent, parallels the former
reference, but we do not attempt any comparison at this stage, as
the formalisms and contexts are entirely different.

In Sec. II we derive the BCS equations, which include the effects
of the finite momentum of the Cooper pairs, within the finite
temperature real-time Green's functions formalism.
The numerical solutions of these equations are shown in Sec. III.
Sec. IV contains a summary of the results.

\section{Formalism}
Below, we shall use the real-time Green's functions extended to
the Nambu-Gor'kov space to account for pair correlations. The single
particle retarded Green's function in this space is defined as
usual:
\be
i\hat G_{\alpha\beta}(x_1,x_2) \equiv i\left( \begin{array}{cc}
        G_{\alpha\beta}(x_1,x_2) & F_{\alpha\beta}(x_1,x_2) \\
        F_{\alpha\beta}^{\dagger}(x_1,x_2)
        & G_{\alpha\beta}^{\dagger}(x_1,x_2)
      \end{array} \right)
=\left( \begin{array}{cc}
\left <T_t\psi_{\alpha}(x_1)  \psi_{\beta}^{\dagger}(x_2)\right >
& \left <T_t\psi_{\alpha}(x_1)\psi_{\beta} (x_2)\right > \\
\left <T_t\psi_{\alpha}^{\dagger}(x_1)\psi_{\beta}^{\dagger}(x_2)\right >
& \left <T_t\psi^{\dagger}_{\beta}(x_1)\psi_{\alpha}(x_2)\right >
      \end{array} \right),
\ee
where $\alpha$ and $\beta$ stand for discrete quantum numbers (spin,
isospin, etc.), $\psi^{\dagger}$ and $\psi$ are the nucleon creation
and annihilation operators, $x\equiv (\br,t) $
denotes the space-time coordinate,
and $T_t$ is the time-ordering symbol.
The averaging is carried out over
the equilibrium ensemble at a fixed density and temperature.
The equation of motion for the matrix Green's function is given by the
time-dependent Dyson equation
\be
\hat G_{\alpha}^{-1}(x_1)\hat G_{\alpha\beta}(x_1,x_2)
= \hat {\bf 1} \delta_{\alpha\beta}\delta (x_1-x_2)
+i\sum_{\gamma}\int d^3x_3~\hat\Sigma_{\alpha\gamma} (x_1,x_3)
\hat G_{\gamma\beta}(x_3,x_2),
\ee
where $\hat {\bf 1}$ is a unit matrix in the Nambu-Gor'kov space,
the inverse free-particle propagator and the self-energy matrixes are
\be
\hat G_{\alpha}^{-1}(x) = \left( \begin{array}{cc}
        G_{\alpha}^{-1}(x) &0 \\
        0 & \left[ G_{\alpha}^{-1}(x) \right]^*
      \end{array} \right),
\quad
\hat \Sigma_{\alpha\beta} (x_1,x_2)\equiv \left( \begin{array}{cc}
        \Sigma_{\alpha\beta} (x_1,x_2) &\Delta_{\alpha\beta}(x_1,x_2) \\
        \Delta_{\alpha\beta} ^{\dagger}(x_1,x_2)
         & \Sigma_{\alpha\beta} ^{\dagger}(x_1,x_2)
      \end{array} \right),
\ee
where $G_{\alpha}^{-1}(x)\equiv  i \partial/\partial t
        +{\nablavec^2}/{2m_{\alpha}} + \mu_{\alpha}$,
$\mu_{\alpha}$ is the chemical potential,
and $m_{\alpha}$ is the bare mass.
The self-energy matrix is defined according to the rules of the
usual diagram technique in the terms of $\hat G$ and the four-fermion
interaction vertex $\hat\Gamma$. In particular, the anomalous self-energy,
which incorporates the pair correlations, is given by
\be\label{GAP}
\Delta_{\alpha\beta}(x_1,x_2) =\sum_{\gamma\kappa}
\int \Gamma_{\alpha\beta\gamma\kappa}(x_1,x_2;x_3,x_4)
F_{\gamma\kappa}(x_3,x_4)dx_3dx_4 .
\ee
In the following we shall be interested in stationary (time-independent)
and spatially inhomogeneous solutions of the equations above in the
quasiclassical approximation. This approximation holds when
the characteristic length scales
of the spatial variations of the macroscopic condensate are much larger
than the inverse of the momenta involved in the problem $\sim p_F$, where
$p_F$ is the Fermi momentum. The quasiclassical
counterparts of the equations above are obtained by going over to the
center of mass $X = (x_1+x_2)/2$ and relative $x = x_1-x_2$ coordinates in
the two-point functions and carrying a Fourier transform with respect
to the relative coordinates:
$\hat G(x, X)\to \hat G(\omega ,\bp, \bR, T)$, where $\omega, \bp$ are
the relative frequency and momentum, and $X \equiv (\bR, T)$.  As the
variations of the propagators and self-energies are slow on the scales
of the order of $\bR$, keeping the leading order terms in the
gradient expansion is accurate to order $\sim {\cal O}[(p_FR)^{-1}$].
Carrying out a Fourier transformation with respect to  $\bR$,
we arrive at the Dyson equation for the quasiclassical functions:
\be\label{QC}
\sum_{\gamma}\left( \begin{array}{cc}
\omega - \epsilon_{\alpha\gamma}^+
       &-\Delta_{\alpha\gamma} \\
       - \Delta_{\alpha\gamma} ^{\dagger}
& \omega + \epsilon_{\alpha\gamma}^-
      \end{array} \right) \left( \begin{array}{cc}
        G_{\gamma\beta}& F_{\gamma\beta} \\
        F_{\gamma\beta}^{\dagger}
        & G_{\gamma\beta}^{\dagger}
      \end{array} \right)= \delta_{\alpha\beta} \hat {\bf 1},
\ee
where
\be
\epsilon_{\alpha\beta}^{\pm}=
 \left(\bP/2\pm\bp\right)^2/2m_{\alpha}-\mu_{\alpha}
\pm {\rm Re~}\Sigma_{\alpha\beta} -{\rm Im~}\Sigma_{\alpha\beta},
\ee
and ${\rm Re~}\Sigma_{\alpha\beta} \equiv
(\Sigma_{\alpha\beta}-\Sigma_{\alpha\beta}^{\dagger})/2$,
${\rm Im~}\Sigma_{\alpha\beta} \equiv
(\Sigma_{\alpha\beta}+\Sigma_{\alpha\beta}^{\dagger})/2$;
all propagators and self-energies are functions of
$\omega, \bp$ and $\bP$ (the dependence on center-of-mass
time is dropped in the stationary limit). Equation (\ref{QC}) is a
$(4\times 4)$ matrix in the spin-isospin space in general.
The number of degrees of freedom can be reduced since
the fermionic wave function of paired fermions must be antisymmetric.
In the case of spin and isospin conserving forces the normal
Green's functions and self-energies are diagonal in the spin and
isospin spaces. It is sufficient to consider the anomalous
propagators, e.g., in the isospin space, since the resulting spin structure,
for $S$ wave interactions, is uniquely determined for each isospin
combination. The quasiparticle excitation spectrum is determined
in the standard fashion by finding the poles of the propagators
in Eq. (\ref{QC}):
\be\label{SPECTRUM}
\omega_{\pm\pm} = \epsilon_A\pm\sqrt{
\epsilon_S+\frac{1}{2}\Tr (\Deltavec\Deltavec^{\dagger})
\pm\frac{1}{2}\sqrt{[\Tr(\Deltavec\Deltavec^{\dagger})]^2 -
4{\rm Det}\left(\Deltavec\Deltavec^{\dagger}\right)
}}.
\ee
Here $\Deltavec\equiv \Delta_{\alpha\beta}$,
$\epsilon_S = (\epsilon^{+}+\epsilon^{-})/2$, and
$\epsilon_A = (\epsilon^{+}-\epsilon^{-})/2$.
The new quasiparticle spectrum has four branches.
The fourfold splitting of the  BCS spectrum is due to
(a) isospin asymmetry and/or the finite
momentum of the condensate
and/or the finite lifetime of the quasiparticles;
and (b) the coupling of the pairing gaps in different
isospin channels. If we restrict ourselves to the neutron-proton pairing in
the $^3S_1$-$^3D_1$ channel, which is justified when
$\Delta_{nn},\Delta_{pp} \ll \Delta_{np}$,
then $\Delta_{\alpha\beta}=\sigma_x\Delta$ ($\sigma_x$ is the first
component of the vector of Pauli matrixes).
The spectrum, in this case, simplifies to
\be\label{SPECTRUM2}
\omega_{\pm} = \epsilon_A \pm \sqrt{\epsilon_S^2+\vert \Delta\vert^2},
\ee
where the symmetric and asymmetric
parts of the spectrum (which are even and odd
with respect to the time-reversal symmetry) are defined as
\be
\epsilon_S\equiv P^2/8m+p^2/2m+{\rm Re~}\Sigma -\mu ,
\quad \epsilon_A\equiv\bP\cdot \bp/2m
+{\rm Im~}\Sigma -\delta\mu.
\ee
Here $\mu =( \mu_n+\mu_p)/2$, $\delta\mu = (\mu_n-\mu_p)/2$ and
${\rm Re~}\Sigma \equiv (\Sigma_{nn} - \Sigma_{pp}^{\dagger})/2$,
${\rm Im~}\Sigma\equiv(\Sigma_{nn} + \Sigma_{pp}^{\dagger})/2$
(subscripts $n$ and $p$ refer to protons and neutrons).
The limit $\epsilon_A \to 0$ corresponds to the BCS pairing
in the isospin symmetric nuclear matter. It is explicit now
that the spectrum (\ref{SPECTRUM2}) is twofold split
due to three factors, the isospin
asymmetry ($\delta\mu \neq 0)$, the finite-momentum of the
Cooper pair ($\bP\neq 0$), and  the finite lifetime
of the quasiparticle  (${\rm Im}\Sigma \neq 0)$. Below, we shall
keep the first two factors, and shall neglect the last one, since
the quasiparticle approximation is valid in the density-temperature
range of interest (densities around the
nuclear saturation density and temperatures $\sim$ 10 MeV).
The solution of the Dyson equation (\ref{QC}) is
\be
G_{n/p} &=& \frac{u_p^2}{\omega-\omega_{+/-}+i\eta}
  +  \frac{v_p^2}{\omega-\omega_{-/+}+i\eta}, \\
F &=& u_p v_p \left( \frac{1}{\omega-\omega_{+}+i\eta}
      -\frac{1}{\omega-\omega_{-}+i\eta}\right),
\ee
where the Bogolyubov amplitudes are
\be
u_p^2 = \frac{1}{2} + \frac{\epsilon_S}
          {2\sqrt{\epsilon_S^2+\vert \Delta\vert^2}} , \quad
v_p^2 = \frac{1}{2} - \frac{\epsilon_S}
          {2\sqrt{\epsilon_S^2+\vert \Delta\vert^2}} .
\ee
Let us turn to the solution of the gap equation (\ref{GAP}),
which in the quasiclassical limit takes the form
\be\label{GAP2}
\Delta (\bp,\bP) = 2\int \frac{d\omega' d^3p'}{(2\pi)^4}
\Gamma(\omega',\bp,\bp',\bP)
{\rm Im} F (\omega',\bp',\bP) f(\omega'),
\ee
where $f(\omega)=\left[{\rm exp}(\beta\omega)+1\right]^{-1}$is the Fermi
distribution function and $\beta$ is the inverse temperature; the
effective pairing interaction $\Gamma$ is assumed real and will be
replaced by the bare interaction below.
The $\omega$ integration is straightforward in the quasiparticle
approximation, since the frequency dependence of the propagator
is constrained by the on-shell condition.
Further progress requires partial wave
decomposition of the interaction, which can be done after
an angle averaging in the remainder functions on the
right-hand side of the Eq. (\ref{GAP2}).
The result of this procedure is
\be\label{GAP3}
\Delta_l(p,P)=-\sum_{l'} \int \frac{dp' p'^2}{(2\pi)^2} V_{ll'}(p,p')
\frac{\Delta_{l'}(p',P)}{2\sqrt{\epsilon_S^2+ \Delta(p',P)^2}}
\langle \left[f(\omega_+)-f(\omega_-)\right]\rangle ,
\ee
where  $\langle \dots \rangle $ denotes the average over the angle between
the relative and total momenta, and
 $\Delta(p,P)^2 \equiv \Delta_0(p,P)^2
+ \Delta_2(p,P)^2$ is the angle-averaged gap. Here the pairing
interaction is approximated
by the bare neutron-proton interaction $V(\bp,\bp') $
in the $^3S_1$-$^3D_1$-channel.
The self-consistent procedure of the determination of the gap function
requires a normalization condition
for the net density $\rho\equiv \rho_n+\rho_p$
of the system at a fixed temperature and the
magnitude of the total momentum $P$. The
corresponding expression is provided by
\be\label{DENSITY}
\rho_{n/p}(P)= -2\sum_{\sigma}\int \frac{d^4 p}{(2\pi)^4}{\rm Im}
G_{n/p}(\omega,\bp,\bP) f(\omega) = \sum_{\sigma}\int \frac{d^3p}{(2\pi)^3}
\left\{u_p^2 f(\omega_{\pm})+v_p^2 f(\omega_{\mp})\right\},
\ee
where $\sigma$ stands for quasiparticle spin and
the second equality follows in the quasiparticle approximation.
The coupled equations (\ref{GAP3}) and (\ref{DENSITY})
should be solved simultaneously.

To find the true ground state we have to
minimize the free energy of the system
at fixed total density and temperature.
In the mean-field approximation  the entropy
of the system is given by the combinatorical expression
\be\label{ENTROPY}
S = - 2 k_B \sum
\Bigl\{f(\omega_{+})\,{\rm ln}\, f(\omega_{+})+
        \bar f(\omega_{+})\,{\rm ln}\,\bar f(\omega_{+})
+f(\omega_{-})\,{\rm ln}\, f(\omega_{-})
+\bar f(\omega_{-})\,{\rm ln}\, \bar f(\omega_{-})
\Bigr\},
\ee
where $\bar f(\omega_{\pm}) = [1-f(\omega_{\pm})]$, and
$k_B$ is the Boltzmann constant.
The internal energy, defined as the grand canonical
statistical average of the Hamiltonian,
$U = \langle \hat H-\mu^{(n)}
\hat\rho_n - \mu^{(p)}\hat\rho_p\rangle$, reads
\be\label{U}
U = 2\int \frac{d^3p}{(2\pi)^3}\Biggl\{
\left[\epsilon^+n_n(p)+\epsilon^- n_p(p) \right]
+\sum_{ll'}\frac{d^3p'}{(2\pi)^3}
\, V_{ll'}(p,p')\, \nu_l(p)\nu_{l'}(p')\Biggl\},
\ee
where
\be
n_{n/p}(p)\equiv u_p^2 f(\omega_{\pm})+v_p^2 f(\omega_{\mp}) ,
\quad
\nu(p) \equiv u_pv_p\langle \left[f(\omega_+)-f(\omega_-)\right]\rangle .
\ee
The first term in Eq. (\ref{U}) includes the kinetic and non-pairing
energies of the quasiparticles.
The second term includes the BCS mean-field
interaction among the particles in the condensate.
The free energy (at fixed density and temperature) is defined as
\be\label{GR}
(F)_{\rho,\beta} =  U - \beta^{-1}S.
\ee
The true ground state of the system minimizes the free-energy
difference  $(\delta F)_{\rho,\beta}$ between the superconducting
and normal states [the free energy in the normal state
follows from Eqs. (\ref{ENTROPY}) and
(\ref{U}) when $\Delta= 0$].

\section{Results}

The main focus of  the numerical calculations
shown below is the effects of the finite
momentum of the Cooper pairs and the emergence of the LOFF
phase in the asymmetric nuclear matter. A number of
simplifying assumptions went into these calculations: first,
the pairing interaction is approximated by the bare interaction;
i.e., the effects of the screening of the pairing interaction
are ignored. Second, we employ
the quasiparticle approximation and set the effective mass
of the quasiparticles equal to their bare mass. Third, we ignore the
coupling  between the pairing in the isospin
triplet and singlet states. The first two approximations change
the absolute magnitude of the paring gap by affecting, respectively,
the strength of the interaction and the density of states at the
Fermi surface. To remove the dependence on the absolute scale
of the gap we present the results normalized to the pairing
gap in the symmetric matter at zero total momentum of the pairs.
The third approximation is valid whenever the pairing
in the isospin singlet channel is much larger than in the
isospin triplet channel.
This could be the case since the strength of the
interaction in the $^3S_1$-$^3D_1$ in the free space
is much larger the one in the  $^1S_0$ channel
and these channels are attractive in the same range of the
energies. This argument, however, implicitly assumes that effects of
the quasiparticle renormalization and the screening of the
pairing interaction are of the same order in both channels,
which could be false.

Figure 1 shows the pairing gap $\Delta(p_F)$ in the $^3S_1$-$^3D_1$
partial wave channel as a function
of the isospin asymmetry, defined as $\alpha \equiv (\rho_n-\rho_p)/\rho$
and total momentum $P$ in units of the Fermi momentum.
The pairing interaction has been approximated by a separable
form of the Paris nucleon-nucleon interaction (PEST1 of Ref. \cite{HP}).
The pairing gap is normalized to its value in the symmetric
and zero-total-momentum case $\Delta_{00} = 14.38$ MeV at
fixed total density  $\rho = 0.16$ fm$^{-3}$ and the temperature
$\beta^{-1}=3 $ MeV. The results reported here are relevant for the
low-temperature regime ($\beta^{-1}/\Delta_{00}=0.208 \ll 1$);
the temperature dependence of the LOFF phase, in particular
the $T\to T_c$ limit, will be discussed elsewhere.

The absolute magnitude of the gap is consistent with the previous
results based on the free-single-particle spectrum\cite{SAL}
(although, note that the gap in Ref. \cite{SAL} is by 15$\%$ smaller,
since there a rank 4 potential has been used  instead of the rank 1
potential used in this work).  A renormalization of the single
particle spectrum, for example within the Brueckner theory, leads
to a decrease of the gap by a factor of 2, see Ref. \cite{BALDO}. This
reduction also affects the critical asymmetry at which the BCS
state disappears, by reducing it, e.g., at nuclear saturation
density, from 0.35 for the free-particle spectrum  to 0.11 for the
Brueckner-renormalized particle spectrum \cite{SAL,SL}.
Therefore, the absolute magnitude of the asymmetry, at which the
transition from the BCS to the LOFF phase occurs, and its critical
value, at which the LOFF phase disappears, will be reduced by roughly a
factor of 3, if the renormalization of the single particle spectrum
is carried out within the Brueckner theory.
For $\alpha = 0$ the  gap is maximal at $P=0$, decreases
as the total momentum is increased, and vanishes at the critical
total momentum $P_{c,0}  = 0.558 p_F$.
For $P = 0$ the gap again
decreases as a function of $\alpha$ and vanishes at $\alpha_{c,0}=
0.37$. The onset of the LOFF phase is signaled by the change of the
shape of the constant $\alpha$ slices in the $\alpha$-$P$ plane:
for $\alpha \ge 0.25 $ the maximum of the gap as a function of
$P$ shifts from the $P=0$ to $P\neq 0$ values; i.e., the condensation
energy becomes maximal for $P\neq 0$. The maximum
is located at  $P \simeq 0.3 p_F$ and is not sensitive to the
value of  $\alpha\ge 0.25$. Interestingly, for $\alpha \ge 0.37$
the pairing exists only for finite momentum states; i.e. there is
a nonzero lower critical momentum  at which the pairing disappears.
The maximal critical values at which the pairing disappears in the whole
$\alpha$-$P$ plane are $\alpha_c = 0.41$ and $P_c = 0.3 p_F$.
The main conclusion that can be drawn from the discussion above, is that
two phase transition take place as the isospin asymmetry is increased:
first a phase transition from the BCS superfluid state with $P=0$
to the LOFF superfluid state with $P\neq 0$ and, second, a phase transition
from the LOFF state to the normal (unpaired) state.

Figure 2 displays the latent heat of phase transition
$Q=(S_n-S_s)/\beta$  as a function of the isospin
asymmetry $\alpha$ and total momentum $P$. At the boundary
of the phase transition from superfluid to the normal state
in the $\alpha$-$P$ plane $Q=0$,
$S_s=S_n$;  hence the phase transition is of the second
order (recall that this result holds in the  mean-field
approximation  used in determining the entropy).
In contrast, $Q\neq 0$, for the phase
transition from the BCS to the LOFF phase
and the phase transition is of the first order, except
along the line of the intersection of $Q(\alpha, P)$ surface
with the $\alpha = 0 =P$ plane. Note that this line marks
the region with anomalous negative sign of $Q$ (i.e., in this region
the entropy of the superfluid state is larger than that of
the normal state). This anomaly does not
result in a metastable state, as the net change of the free
energy shown below remains always negative.

Figure 3 shows the difference of the free energies of the
normal and superconducting states  $(\delta F)_{\rho\beta}$,
which is normalized to its value in the symmetric
and zero-total-momentum case $(\delta F_{00})_{\rho\beta}=-7.35$ MeV at
$\rho = 0.16$ fm$^{-3}$ and $\beta^{-1}=3$ MeV.
The onset of the LOFF phase is seen by examining
the constant $\alpha$ slices of the $(\delta F)_{\rho\beta}$
surface. The onset of the LOFF phase is signaled by the change of the
shape of the these curves:
for $\alpha \ge 0.25 $ the minimum of $(\delta F)_{\rho\beta}$
as a function of  $P$ shifts from the  $P=0$ to $P\neq 0$ values;
i.e., the ground state energy corresponds to the state with a total
nonzero momentum of the pairs. The minimum of the free energy difference,
as the maximum of the gap function,
is located at  $P \simeq 0.3 p_F$ and is not sensitive
to the value of  $\alpha\ge 0.25$. The similarity of the
functional dependence of the free energy difference and the
pairing gap on the parameters $\alpha$ and $P$  is not accidental,
as  $(\delta F)_{\rho\beta}$  is dominated by the pair interaction
(condensation) energy given by the second term in Eq. (\ref{U}),
which scales as pairing gap squared.

\section{Summary and Outlook}

In this work we have analyzed the BCS solutions for the neutron-proton
pairing in the asymmetric nuclear matter when the Cooper pairs are
allowed for a nonzero total momentum. The quasiparticle excitation
spectrum is four-fold split compared to the usual BCS spectrum of the
symmetric, homogeneous matter.
The twofold splitting occurs
due to the finite momentum of the condensate and/or the
isospin asymmetry and/or the finite quasiparticle
lifetime; the simultaneous paring in the isospin single and triplet
states leads to a further twofold splitting of the spectrum.
The gap equation, which was solved numerically in the
limiting case of vanishing isospin triplet pairing,
has nontrivial solutions with finite total momentum of the pairs.
The corresponding nuclear LOFF phase is the true ground state
of the system for density asymmetries larger than 0.25. The  minimum of
the free energy corresponds to the total momentum of the
condensate $P = 0.3p_F$ independent of the value of $\alpha$.
For sufficiently large asymmetries ($\alpha \ge 0.3)$ the condensate can
exist only in the state with finite momentum; i.e., apart from a upper critical
total momentum for vanishing of the condensate, there is a lower one at
which condensate sets in. The maximal values of the total momentum and
asymmetry that the condensate can sustain are $P = 0.3 p_F$
and $\alpha = 0.41$. The actual value of $\alpha$  found for
the nonrenormalized
single particle spectrum could be reduced by a
factor of 3 if a renormalization
is carried in the Brueckner theory of nuclear matter.

Thus, in a definite region of the $\alpha$-$P$ plane the neutron-proton
condensation occurs at  nonzero momentum of the
Cooper pairs, which leads to the formation of
a spatially inhomogeneous phase
of nuclear matter. This implies a periodic (translationally and rotationally
invariant with respect to the basis vectors) spatial structure of the
condensate which carries an isospin density wave at constant total number
of particles. One of the consequences of the periodic structure is that
the quasiparticle velocities in certain directions could be close to zero,
which implies a strong anisotropy of the kinetic coefficients of the matter
and larger heat capacity than in the homogeneous phase.

The phase transition from the LOFF phase to the normal (nonsuperconducting)
phase is a  transition of the second order. However, the phase transition
from the BCS to the LOFF phase turns out to be of the first order;
i.e.,
there is a latent heat of transition associated with this phase transition.
In a certain region of the $\alpha$-$P$ plane the latent heat has an
anomalous negative sign.
However, this does not affect the stability of the LOFF
phase, since its energy budget is dominated by the
pair-condensation energy.

What lattice structure prefers the nuclear LOFF phase? The problem of the
energetically most favorable structure of the LOFF phase
has not been solved so far in general. For small gaps the integral
equation (\ref{GAP2}) is linear and we can seek the solutions
in terms of a Fourier expansion
\be
\Delta(\br)  = \sum_{n} \Delta_n e^{i\bP_n \br},
\ee
where the lengths of the ``basis vectors'' $\vert \bP_n\vert $ are
equal. Fulde and Ferrell  studied in detail the thermodynamics
of the LOFF phase with the order parameter containing a single
harmonic: $\Delta(\br)  = \Delta_0 e^{i\bP \br}$ \cite{FULDE}.
Perhaps on symmetry grounds one can argue that a symmetric ansatz
$\Delta (\br) = \Delta (-\br)$, which implies a
real gap function, is the case. In the latter case
the most general form of the harmonic expansion is
\be
\Delta (\br) = 2 \sum_{n} \Delta_n\, {\rm cos} (\bP_n \br).
\ee
The limiting case of a single harmonic $\Delta(\br) = 2
\Delta_0 {\rm cos} (\bP \br)$ has been studied by
Larkin and Ovchinnikov \cite{LARKIN}; in this case one finds a
layered structure. Perhaps, a cubic structure, in which
case  $\Delta(\br) = 2 \Delta_0 \left[{\rm cos} (P x)+
{\rm cos} (P y)+{\rm cos} (P z)\right]$, is preferred to
the layered one if there are no preferred directions
in the problem.
To conclude, the periodic structure of the LOFF phase has been studied only
for limited configurations or spatial dimensions so far. The determination
of the true ground state structure of this phase remains
for the future work.

\section*{Acknowledgments}

I am grateful to  Umberto Lombardo for discussions in the course
of our collaboration on the pairing in nuclear matter and
to Krishna Rajagopal for discussions on the LOFF phase and
for pointing out Ref. \cite{QCDLOFF}. This work has
been  supported by the Stichting voor Fundamenteel Onderzoek der
Materie of the Nederlandse Organisatie voor Wetenschappelijk Onderzoek.

\begin{figure}[t]
\mbox{\epsfig{figure=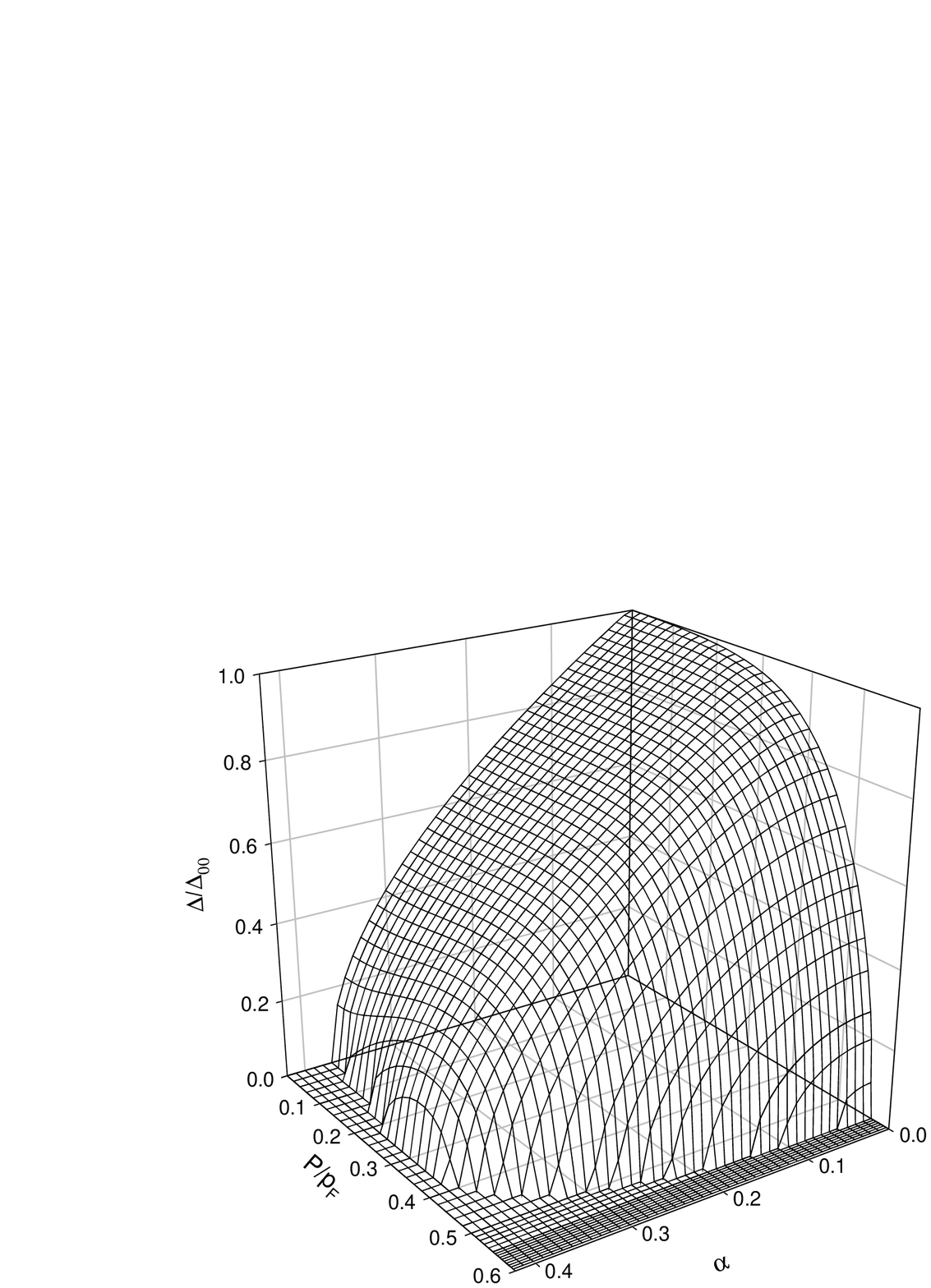,height=7.in,width=5.5in,angle=0}}
\caption[]
{\footnotesize{The pairing gap  $\Delta(p_F)$ in the $^3S_1$-$^3D_1$
partial wave channel  as a function of the isospin asymmetry
$\alpha \equiv (\rho_n-\rho_p)/\rho$ and total momentum $P$
in units of the Fermi momentum.
The pairing gap is normalized to its value in the symmetric
and zero-total-momentum case $\Delta_{00} = 14.38$ MeV at
fixed total density  $\rho = 0.16$ fm$^{-3}$ and temperature
$\beta^{-1} =3$ MeV.}}
\label{fig1}
\end{figure}

\begin{figure}[b]
\mbox{\epsfig{figure=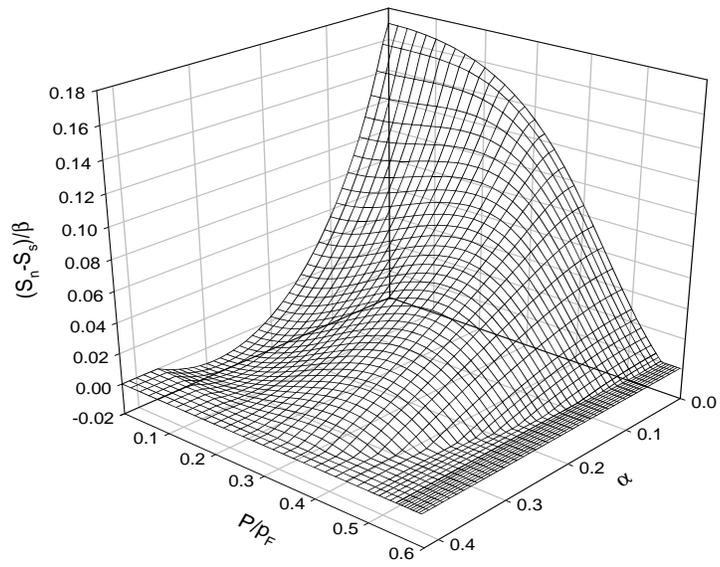,height=7.in,width=5.5in,angle=0}}
\caption[]
{{The latent heat of transition as a function of isospin asymmetry $\alpha$
and total momentum $P$. The remaining parameters are the same as in Fig. 1.}}
\label{fig2}
\end{figure}

\begin{figure}[t]
\mbox{\epsfig{figure=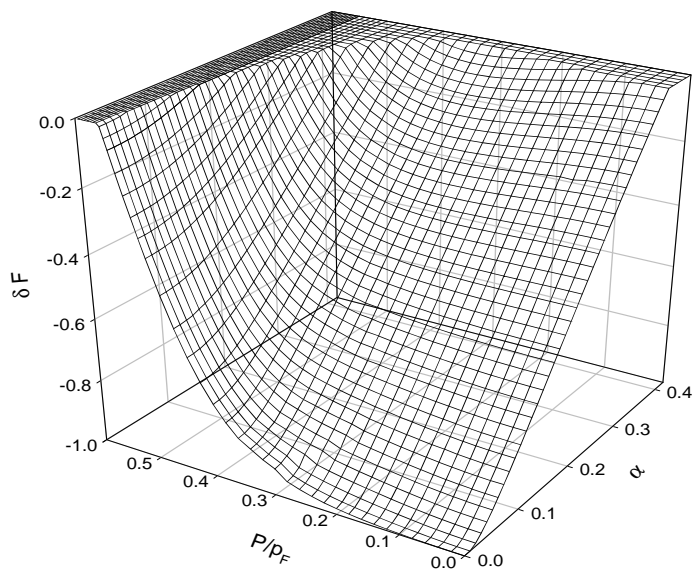,height=7.in,width=5.5in,angle=0}}
\caption[]
{{The difference of the free energies of the normal and superconducting
state  $(\delta F)_{\rho\beta}$,  normalized to its value in the symmetric
and zero-total-momentum state $(\delta F_{00})_{\rho\beta}=-7.35$ MeV.
The remaining parameters are the same as in Fig. 1.}}
\label{fig3}
\end{figure}
\end{document}